# CONTEXT-ENCODING FOR NEURAL NETWORK BASED SKULL STRIPPING IN MAGNETIC RESONANCE IMAGING

*Zhen Liu*[★]*, Borui Xiao*[†]*, Yuemeng Li*[★]*, Yong Fan*[★]

[★]Department of Radiology, Perelman School of Medicine, University of Pennsylvania, Philadelphia, PA, 19104, USA
[†]Department of Biomedical Engineering, School of Biomedical Engineering, Science and Health System, Drexel University, Philadelphia, PA 19104, USA

## ABSTRACT

Skull stripping is usually the first step for most brain analysis process in magnetic resonance images. A lot of deep learning neural network based methods have been developed to achieve higher accuracy. Since the 3D deep learning models suffer from high computational cost and are subject to GPU memory limit challenge, a variety of 2D deep learning methods have been developed. However, existing 2D deep learning methods are not equipped to effectively capture 3D semantic information that is needed to achieve higher accuracy. In this paper, we propose a context-encoding method to empower the 2D network to capture the 3D context information. For the context-encoding method, firstly we encode the 2D features of original 2D network, secondly we encode the sub-volume of 3D MRI images, finally we fuse the encoded 2D features and 3D features with semantic encoding classification loss. To get computational efficiency, although we encode the sub-volume of 3D MRI images instead of building a 3D neural network, extensive experiments on three benchmark Datasets demonstrate our method can achieve superior accuracy to state-of-the-art alternative methods with the dice score 99.6% on NFBS and 99.09 % on LPBA40 and 99.17 % on OASIS.

*Index Terms*— Skull stripping, deep learning, 2D neural network, 3D neural network

## 1. INTRODUCTION

Skull stripping, also called brain extraction, aims to retain brain parenchyema and discard non-brain tissues, such as skull, scalp, and dura. As a fundamental problem in brain MRI image analysis, numerous methods have been proposed over the past 20 years. Some of them are based on morphological operations, e.g., brain surface extraction and some others are based on deformation models that try to fit the brain surface, e.g., the brain extraction tool(BET) [1]. Manual brain extraction is time consuming especially in large-scale studies. Automated brain extraction is necessary but its performance and accuracy are critical as the output of this step can directly affect the performance of all following steps.

Recently neural networks and deep learning have attracted enormous attention in medical image processing. It is the same as to the skull stripping task for brain MRI image analysis. The U-Net [2] architecture is one of the most famous deep learning method for medical image segmentation which stems from the so-called "fully convolutional network (FCN)" [3] first proposed by Long and Shelhamer. The main idea is to supplement a usual contracting network by successive layers, where pooling operations are replaced by upsampling operators. Hence these layers increase the resolution of the output. What's more, a successive convolutional layer can then learn to assemble a precise output based on this information. A 3D CNN architecture [4] uses seven 3D convolutional layers for voxelwise image segmentation. Cubes of size 53 × 53 × 53 around the grayscale target voxel are used as inputs to the network. We refer this work as PCNN in the following parts. In addition to the 3D convolutional network, in [5] a auto-context convolutional neural network has been proposed. Instead of using 3D convolutional layers, 2D patches of three different sizes are adopted and perform convolution with three parallel pathways of 2D convolutional layers in three planes (i.e. axial, coronal and sagittal planes). By adding this context architecture into U-Net, this network is refered as 2.5D-CNN in the following. Besides the U-Net convolutional networks, this architecture also can be adapted with the normal convolutional architecture [5], which is refered as Auto-2.5D-CNN in the following parts.

Besides the deep learning based methods, in this paper, we also compare with the traditional works. The brain extraction tool (BET) [1] is a commonly accepted brain extraction tool. It used a deformable mesh that first initialized at the center of gravity and next, fit the surface of the brain. 3dSkullStrip from the AFNI(Analysis of Functional Neuro Images) package is another software package for brain extraction[6]. Some modifications have been put into practice for BET. Compared to BET, it uses points inside and outside the brain to guide the evolution of the mesh to prevent from segmenting other non-brain tissues, such as the eyes and the skull. Hybrid Watershed Algorithm(HWA) [7] uses watershed algorithm and deformable surface model to do the extraction. The watershed algorithm produces the estimation of brain volume and the deformable surface model produces a force field to lead the sphere to the edge of the brain. Robust learning-Based Brain Extraction(Robex)[8] proposed a discriminative model, which is a random forest classifier for contouring the boundary between brain and non-brain tissue and an active shape model to increase the robustness.

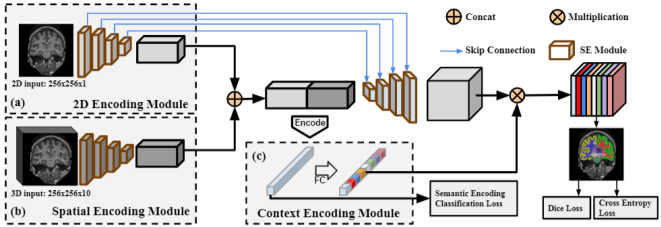

**Fig. 1**. Figure 1: Overview of the proposed network structure. Given inputs of 2D image slice (a) and 3D image volume (b), we first pass each input to a densely-connected encoder to obtain both 2D and 3D intrinsic features. We then build a Context Encoding Module (c) on top of the fused features. The encode module contains an Encoding Layer to capture the encoded global semantic context and predict the scaling factor that highlight the class-dependent variation in module (c) to the encoded semantics.

In this paper, we propose our new CNN architecture inspired by the U-net architecture [2] [9]. We trained and tested our algorithm on three public datasets. The first dataset is publicly available as the Neuro feedback Skull-stripped (NFBS) repository [10]. The second dataset comes from the LONI Probabilistic Brain Atlas Project(LPBA40)[11]. The third dataset is Open Access Series of Imaging Studies(OASIS) [12].

## 2. METHODS

In order to achieve fast and accurate MRI brain extraction, we present our proposed context-encoding architecture, which comprises three encoding modules in Fig. 1. Three encoding model are integrated into a unified model with an end-to-end training procedure: a 2D encoding module for extraction of feature information from the coronal plane, a spatial encoding module that extracts the intrinsic context features and a context encoding module that captures the global semantic contexts. Our network utilizes the densely connected blocks inside the encoder and decoder structure.

### 2.1. Feature Fusion Module

Our proposed feature fusion module is composed of two parts: 2D encoding module in Fig. 1-a and spatial encoding module in Fig. 1-b. The 2D encoding module consists of a set of densely connected blocks, and each followed by a max-pooling block. By taking a single image slice as input, the 2D encoding module effectively extracts dense regional information from the coronal plane. Note that in such a 2D encoding setting, the goal of this module is to effectively generate feature semantics based on the image textures and intensities. Nevertheless, the context features obtained so far are not sufficient for a high accuracy skull stripping. Therefore we utilize the auxiliary 3D spatial information to provide a more comprehensive measurement regarding neuroanatomy, specifically for small size structures such as angular and temporal gyrus etc. Different from conventional 3D network, our proposed spatial encoding module takes the consecutive image slices as input. In this scenario, the third dimension of the input 3D images can be regarded as the H × W × C stacked 2D image slices along the channel, rather than an H × W × D × C depth as in 3D volumes, where H is the height, W is the width, D is the depth, and C is number of channels. By regarding depth as channel from the stacked grayscale image slices (H × W × D) shown in Fig. 1-b, the input for spatial encoding module is constructed in the same dimension as 2D input in Fig. 1-a. This strategy acquires intrinsic spatial context information with less computation compared with a 3D convolution module. This spatial encoding module follows the same structure as the encoder in Fig. 1-a.

### 2.2. Context Encoding Module

The context encoding module is designed in the bottleneck part of the framework in Fig. 1-c. This module consists of an encoding layer, a fully connected layer and an activation function. The encoding layer is incorporated with the anatomical prior to capture the global semantic context. The global semantics obtained from encoding layer is passed through a fully connected layer followed by a sigmoid activation function. The scaling factor $\gamma$ of the class dependent feature maps are predicted from sigmoid function $\sigma(\cdot)$, i.e. $\gamma = \sigma(We)$, where $W$ is the layer weights and $e$ is the encoding output. The network output is calculated as $Y = X \otimes \gamma$, where $\otimes$ is the channel-wise multiplication.

### 2.3. Loss Function

We use three loss functions during network training process: (i) a pixel-wise cross-entropy loss $L_{ce}$, (ii) a multiclass Dice loss $L_{dice}$ and (iii) semantic encoding classification loss $L_{sec}$. The pixel-wise cross-entropy loss provides a similarity comparison between output labels and manual labeled ground truth [13]. Denote $p_l$ as the estimated probability of pixel $x$ belonging to class $l$, and $g_l$ as ground truth labels, the

Table 1. The mean scores of five fold cross validation for different algorithms on NFBS dataset.

| Method | NFBS | | |
|---|---|---|---|
| | Dice | Sensitivity | Specificity |
| Our | 99.6 | 98.53 | 99.73 |

Table 2. The mean scores of two fold cross validation for different algorithms on LPBA40 and OASIS datasets. The results show that our algorithm can obtain the highest dice score and specificity score and comparable sensitivity score.

| Method | LPBA40 | | | OASIS | | |
|---|---|---|---|---|---|---|
| | Dice | Sensitivity | Specificity | Dice | Sensitivity | Specificity |
| FCN | 97.73 | 98.31 | 99.48 | 97.62 | 98.66 | 98.77 |
| U-net | 96.79 | 97.22 | 99.34 | 96.22 | 97.29 | 98.27 |
| Auto-2.5D-CNN | 97.66 | 98.25 | 99.47 | 96.06 | 96.21 | 98.56 |
| 2.5D-CNN | 97.17 | 98.52 | 99.24 | 95.61 | 96.3 | 98.20 |
| PCNN | 96.96 | 97.46 | 99.41 | 95.02 | 92.04 | 99.28 |
| BET | 94.57 | 98.52 | 98.22 | 93.44 | 93.41 | 97.70 |
| Robex | 95.40 | 94.25 | 99.43 | 95.33 | 92.97 | 99.21 |
| 3dSkullStrip | 92.99 | 96.95 | 97.87 | 92.77 | 94.44 | 96.82 |
| HWA | 92.41 | 99.99 | 97.07 | 94.06 | 98.06 | 96.34 |
| Our | 99.09 | 98.45 | 99.84 | 99.17 | 98.61 | 99.82 |
| Our (transfer) | 97.84 | 94.37 | 99.8 | 97.06 | 91.47 | 99.81 |

pixel-wise cross-entropy loss is:

$$L_{ce} = -\sum_{x} \omega(x) g_l(x) log(p_l(x)) \quad (1)$$

. The Dice loss measures the reproduciblity of the model by performing a pair-wise comparison between the generated label and ground truth. We adopt the the multi-class Dice loss setting from QuickNAT [14], which is formulated as:

$$L_{dice} = -\frac{2\sum_x p_l(x) g_l(x)}{\sum_x p_l^2(x) + \sum_x g_l^2(x)} \quad (2)$$

. The standard training per-pixel loss focuses on each individual pixel, while the global information represented by each class label are often overlooked. We therefore utilize a Semantic Encoding Classification Loss (SEC-loss) to force network to focus on finding the corresponding global semantics from the given class label:

$$L_{sec} = -\frac{1}{C} \sum_{i=1}^{C} y_i \cdot log(p(y_i)) + (1-y_i) \cdot log(1-p(y_i)) \quad (3)$$

where C as number of classes, $\gamma$ is the ground truth class label, and $p_y$ is the predicted probability of the class. This loss is applied on the global automatic semantics from the context encoding module as shown in Fig. 1-c. To this end, the overall loss could be formulated as:

$$L_{total} = L_{ce} + L_{dice} + \lambda L_{sec} \quad (4)$$

where $\lambda = 0.1$ is a weight balancing factor.

## 3. EXPERIMENTAL RESULTS

### 3.1. Datasets

We evaluated our algorithm first on two publicly available benchmark datasets and then perform the transfer evaluation with the learned model on another dataset. Two-fold cross-validation is used in all experiments. The output of all algorithms was evaluated against the ground truth which was available for the benchmark datasets. The first dataset is a database of 125 T1-weighted anatomical MRI scans that are manually skull-stripped which is publicly available as the Neurofeedback Skull-stripped (NFBS) repository [10]. The second dataset came from the LONI Probabilistic Brain Atlas Project (LPBA40)[15]. This dataset consists of 40 T1-weighted MRI scans of healthy subjects with spatial resolution of 0.86 × 1.5 × 0.86 mm. The third dataset involved the fist two disks of the Open Acess Series of Imaging Studies (OASIS) [12]. This consisted of 771 × 1 × 1 mm T1-weighted MRI scans of healthy subjects and subjects with Alzheimer's disease.

### 3.2. Implementation Details

Our network takes 2D image slices of 256 × 256 and a 3D image volume of 256 × 256 × 10 as inputs. Both inputs are selected in a coronal view. We employed the learning rate scheduling "poly" from [16], with a start learning rate of 0.01 for coarse structures and 0.02 for fine-grained structures. The weight decay rate is set to $1 \times 104$ for all tested models. A dropout rate of 0.1 is applied to each densely con-nected block. The model is trained for 100 epochs in total. All experiments are performed on a single NVIDIA TITAN XP GPU with 12GB of RAM.

### 3.3. Results

In Table 1, we can get the very high dice score with 99.6 % with five fold cross validation. As shown in Table 2, our context-encoding method can obtain significant improvement for the dice score while keeping the comparable sensitivity and specificity score comparing to the existing algorithms on both LPBA40 and OASIS datasets. From the above data, although the proposed context-encoding CNN is a 2D network architecture, it is clearly that we can get much higher accuracy than the 3D covolutional networks, namely PCNN [4] which improved from 96.96% to 99.09 % on LPBA40 and from 95.02 % to 99.17 % on OASIS. Even for the recently proposed work [5], which also use context information in 2D network architecture, the dice score can also be improved from 97.73 % to 99.09 % and from 97.62 % to 99.17 % respectively. When we use the trained model on NFBS dataset of Table 1 and perform testing on LPBA40 and OASIS namely transfer, the dice score of 97.84 % and 97.06 % can be achieved. It can be seen that the transfer result is still high and comparable to the non transfer result.

## 4. CONCLUSION

In this paper, we introduced an context-encoding model for MRI skull stripping. Our model learns contexts information through a global anatomical prior and utilizes the 3D spatial information to achieve the fast and accurate volumetric skull stripping. We transformed our feature fusion module into a 2D encoding module and a spatial encoding module to extract both regional and spatial features. We proposed to use a semantic encoding classification loss to enforce the learning of global information from the input slices. We tested our model on two datasets, and achieved prominent improvement compared with the state-of-the-art approaches.